\documentclass{webofc}

\usepackage[varg]{txfonts}   
\usepackage{hyperref}
\usepackage{url}
\hypersetup{colorlinks=true,citecolor=blue,urlcolor=blue,linkcolor=blue}
\begin{document}
\title{SRF programs towards High-Q/High-G cavities in IJCLab}
%
%

\author{\firstname{Akira} \lastname{Miyazaki}\inst{1}\fnsep\thanks{\email{Akira.Miyazaki@ijclab.in2p3.fr}} \and
        \firstname{Mohammed} \lastname{Fouaidy}\inst{1}\fnsep \and
        \firstname{Nicolas} \lastname{Gandolfo}\inst{1}\fnsep \and
        \firstname{David} \lastname{Longuevergne}\inst{1}\fnsep \and
        \firstname{Guillaume} \lastname{Olry}\inst{1}\fnsep \and
        \firstname{Mael} \lastname{Vannson}\inst{1}\fnsep \and
        \firstname{L\^{e} My} \lastname{Vogt}\inst{1}\fnsep \and
        \firstname{Matthieu} \lastname{Baudrier}\inst{2}\fnsep \and
        \firstname{Enrico} \lastname{Cenni}\inst{2}\fnsep \and
        \firstname{Fabien} \lastname{Eoz\'{e}nou}\inst{2}\fnsep \and
        \firstname{Gr\'{e}goire} \lastname{Jullien}\inst{2}\fnsep \and
        \firstname{Luc} \lastname{Maurice}\inst{2}\fnsep
}

\institute{CNRS/IN2P3/IJCLab Universit\'{e} Paris-Saclay
\and
           CEA/IRFU/DACM Universit\'{e} Paris-Saclay
          }

\abstract{
IJCLab has been leading the development and deployment of low-$\beta$ Superconducting Radio Frequency (SRF) cavities for proton and heavy ion accelerators.
We are launching an electron accelerator project for sustainable Energy Recovery Linac (iSAS/PERLE) with state-of-the-art SRF cavities at 800MHz.
Our proposal includes advanced heat treatment of such cavities to reach an excellent quality factor of $3\times 10^{10}$ at $22$~MV/m.
In this paper, we overview the status of this activity.
}
\maketitle
\section{Previous projects at IJCLab} \label{sec:intro}
The accelerator division in Le laboratoire de physique des deux infinis Ir\`{e}ne Joliot-Curie (IJCLab) is fusion of accelerator departments of former 
le laboratoire de l'acc\'{e}l\'{e}rateur lin\'{e}aire (LAL)
and
l'institut de physique nucl\'{e}aire d'Orsay (IPNOrsay).
Regarding the Superconducting Radio Frequency (SRF) cavities for particle accelerators,
we have been leading the development of low-$\beta$ cavities for protons (ESS, MYRRHA, PIP-II) and heavy ions (Spiral2) accelerators~\cite{longuevergne18, PhysRevAccelBeams.24.083101, Miyazaki:2023zye}.
These cavities include 88MHz quarter-wave resonators, 325MHz single-spoke cavities, and 352MHz single- and double-spoke cavities.
For $\beta \sim 1$ electron accelerators, we developed and provided more than 800 high-power RF couplers to the Eu-XFEL project~\cite{Kaabi:2013wna}
together with contributions to beam dynamics calculations in various accelerators.
The SRF cavities for electrons have not been the major focus of the activities at IJCLab.
IJCLab has started to develop an on-site electron accelerator with SRF cavities.

\section{New projects: iSAS/PERLE} \label{sec:new}
For future accelerators, one of the pressing issues is the sustainability of accelerator projects, in particular, electric power consumption in the context of recent energy crisis and global warming.
For instance, future energy frontier colliders for Higgs bosons and new physics beyond the Standard Model are subject to a significant challenge in their sustainability.
One potential solution to reduce both electricity and construction costs for the Higgs factory is the Large Hadron electron Collider (LHeC)~\cite{Abelleira_Fernandez_2012} 
that includes a new electron accelerator at CERN to be combined with the existing Large Hadron Collider (LHC).
The electron-proton collisions at high energy generate Higgs bosons mainly via Vector Boson Fusion (VBF) and are complementary to hadron-hadron collider (HL-LHC) at 14~TeV and electron-positron colliders (FCCee, CEPC, ILC, CLIC etc) below 500~GeV.
The key idea of LHeC is to introduce Energy Recovery Linac (ERL) technology for the electron accelerator.
The ERL concept, which combines the high beam quality of linear accelerators and the high current of circular accelerators, allows to achieving high luminosity in an energy-efficient way.
In principle, ERLs inject electron bunches back into the accelerating cavities in the decelerating phase after their use in experiments, such as collisions to another beam, free electron laser etc.
Since the electromagnetic interaction is reversible, ideally, the energy of the beam is transferred to the cavities and is used to accelerate the next bunches of low emittance.

A challenge of ERLs is technical readiness at a high current of more than 10~mA.
In IJCLab, we are constructing a demonstrator of the ERL for LHeC, Powerful Energy Recovery Linac for Experiments (PERLE), 
with the same single-turn current of 20~mA as LHeC and beam energy of 500~MeV.
In parallel to the construction of the PERLE accelerator, 
IJCLab is coordinating a more general European project, innovate for Sustainable Accelerating System (iSAS) towards the realization of energy saving in accelerator technology.
In these two collaborations, we are developing new technology for the post HL-LHC era.

IJCLab develops an iSAS/PERLE cryomodule (CM) that contains four SRF cavities.
These SRF cavities have a 5-cell $\beta=1$ elliptical shape with a frequency of 801.580MHz, the 2nd harmonics of the existing LHC RF frequency.
The cavities are made of high-purity (residual resistivity ratio 300) bulk niobium as a common technical choice in state-of-the-art SRF accelerators.
To achieve the electron beam energy with reasonable cryogenic power consumption below 2~K in Continuous Wave (CW) RF operation,
the iSAS/PERLE project adopted an extremely high unloaded quality factor of $Q_0=3\times10^{10}$ at a reasonably high accelerating gradient of $E_{\rm acc}=22$~MV/m.
Prototype cavities of PERLE were designed and fabricated by Jefferson Laboratory (JLAB) in the United States in the common framework of FCC prototyping in 2018~\cite{Marhauser:2653853}.
The surface treatment included bulk Buffered Chemical Polishing (BCP) of 200~${\rm \mu}$m, 
hydrogen degassing at 800$^\circ$C for 3~hours, 
flush Electropolishing (EP) of 30~${\rm \mu}$m,
and baking at 120$^\circ$C for 48~hours.
The 5-cell prototype met the specification in a cold test at a vertical test stand at JLAB.
Despite this successful implementation,
it is known that cavities in a CM are often degraded by 10-15\% from the vertical test;
thus, a sufficient safety margin is of critical importance in cavity development.
The next challenge of iSAS/PERLE cavities is to obtain their ambitious specification inside the CM.
For this purpose, we are currently working on advanced heat treatment~\cite{DHAKAL2020100034} for excellent $Q_0$.

\section{Advanced heat treatment} \label{sec:HH}
It has been known that baking at relatively low-temperatures around 120$^\circ$C can improve the maximum accelerating gradient by eliminating non-linear surface resistance called high-field-Q-slope~\cite{Steder:2019nrv}.
In the ILC standard recipe of cavity surface preparation, which was also adopted for European XFEL, EP followed by low temperature baking (low-T bake) is performed.
Some variants of low-T bake, such as two-step baking, have been investigated for better production yield of high-gradient machines.

For high-current and CW machine such as PERLE, a higher quality factor is more critical than a higher accelerating gradient, to minimize cryogenic power consumption.
One promising way to achieve a higher quality factor around 20~MV/m is to inject nitrogen during the annealing process above 800$^{\circ}$C for 2 minutes.
This process is called nitrogen doping (N-dope) and was adopted to LCLS-II cavities and high-$\beta$ cavities of PIP-II~\cite{GONNELLA2018143}.
One drawback of N-dope is that it requires uniform EP after the treatment to remove unwanted nitride layers;
thus, the $Q_0$ improvement is supposed to be caused by the remaining interstitial nitrogen inside the niobium material.

Another process to dramatically improve $Q_0$ without nitrogen is to simply bake the niobium material around 300$^\circ$C for a few hours~\cite{PhysRevApplied.13.014024}.
Importantly, the cavity's inner surface must be exposed to air before this treatment.
This alternative heat treatment is called mid-T bake and is gaining popularity thanks to its simple procedure and promising results.
The most plausible theoretical model is that mid-T bake diffuses oxygen from the natural surface oxide layer,
which is inevitable when niobium is exposed to air.
Interstitial oxygen modifies the scattering processes of quasi-particles and Cooper pairs inside the superconductors.
In this sense, mid-T bake can be considered as oxygen doping (O-dope).
An excellent $Q_0$ has been obtained in 650MHz cavities~\cite{app12020546}.
IJCLab will optimize the mid-T bake for the iSAS/PERLE project with the industry in collaboration with other laboratories in the world.

\section{First mid-T bake at IJCLab} \label{sec:mid-T}
To initiate the mid-T bake activities of iSAS/PERLE cavities in IJCLab,
we performed heat treatment with a smaller prototype cavity at 1.3GHz.
This is a cavity (1DE8) prepared by DESY in the 1990's and is currently on loan for R\&D projects at IJCLab.
At the SUPRATECH platform~\cite{supratech},
IJCLab owns a clean vacuum furnace with cryogenic pumping~\cite{8328862}, 
which has been used for 650$^\circ$C annealing of more than 26 double-spoke cavities for the ESS project to degas hydrogen after BCP~\cite{9366923}.
The unique point of this spoke cavity treatment was that the cavities were heat-treated with a helium jacket made of titanium.
Other facilities usually perform heat treatment with bare niobium cavities without titanium.
We achieved excellent performance of spoke cavities in the ESS project~\cite{Miyazaki:2023zye}.
The use of the same furnace for the mid-T bake process is a new challenge in IJCLab.
\begin{figure}[h]
\centering
\includegraphics[width=12cm,clip]{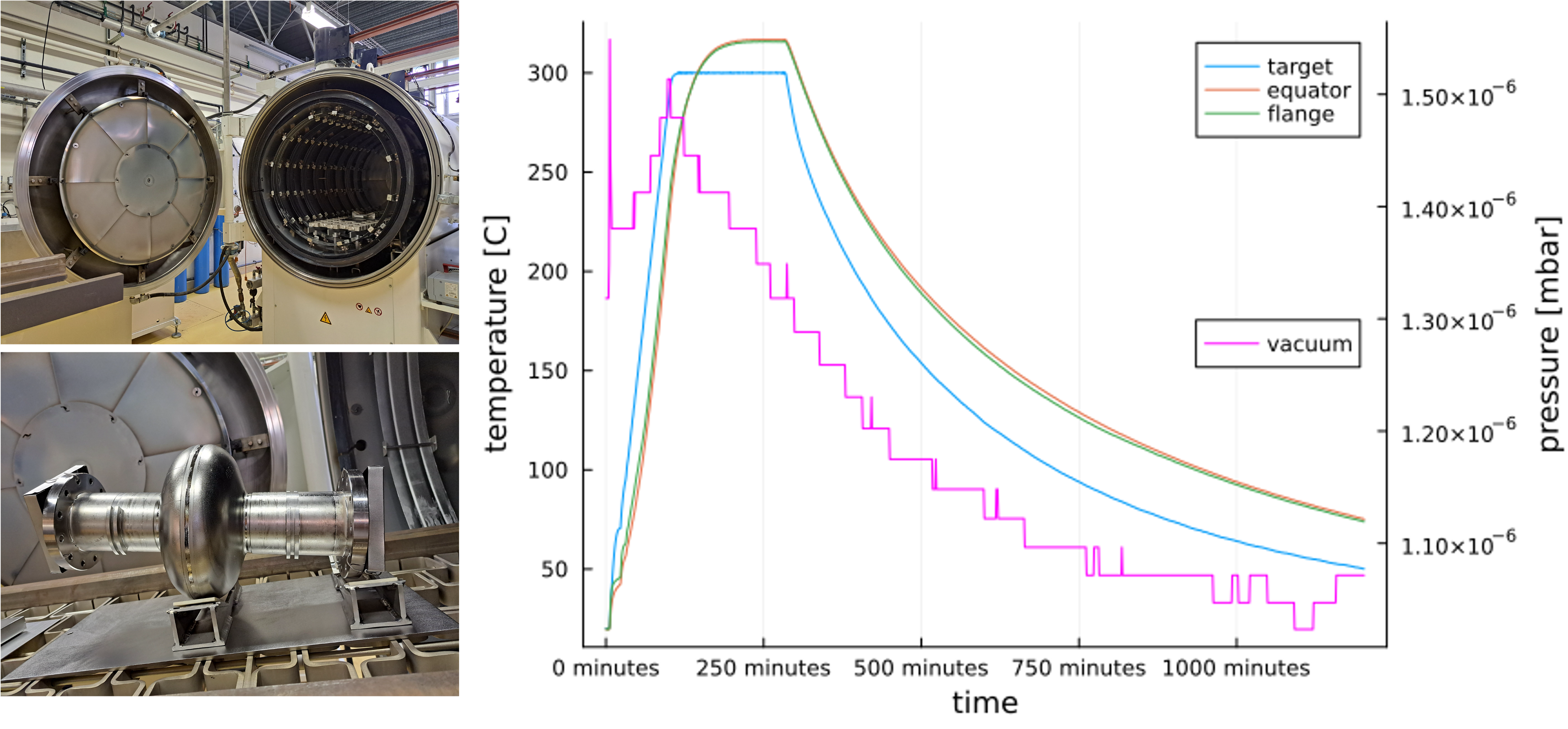}
\caption{First mid-T bake at IJCLab. The top left shows the vacuum furnace. The bottom left shows the treated cavity. The right shows the temperature and vacuum pressure during the baking process.}
\label{fig:mid-T-photo}       
\end{figure}

The first goal of mid-T bake is to see the potential contamination level of the furnace.
In March 2024, we performed {\it dry} annealing of the furnace at 800$^\circ$C for three hours to investigate and/or remove potential contamination.
In April 2024, we did the first mid-T bake of the 1.3GHz cavity as shown in Fig.~\ref{fig:mid-T-photo}.
The cavity was prepared in our ISO4 clean room after high-pressure water rinsing.
The beam pipe was covered by lids made of niobium to avoid contamination during the heating process.
Since our furnace is not inside a clean room, special care was taken during the transport.
The cavity was packed in a plastic bag inside the clean room and then installed in the furnace immediately after the bag was removed, 
minimizing the time when the cavity was exposed to the potentially contaminated air around the vacuum furnace.
The vacuum reached around $10^{-6}$~mbar with the cryogenic pump before the heating process.
The furnace reached 300$^\circ$C within 2 hours and kept its temperature at 300$^\circ$C for 3 hours as the system was pre-programmed.
We observed 16$^\circ$C offset on the cavity temperature compared to the target temperature at 300$^\circ$C.
After the heating process, the furnace was cooled down naturally and finally vented with ultra-pure argon gas.

\begin{figure}[h]
\centering
\includegraphics[width=10cm,clip]{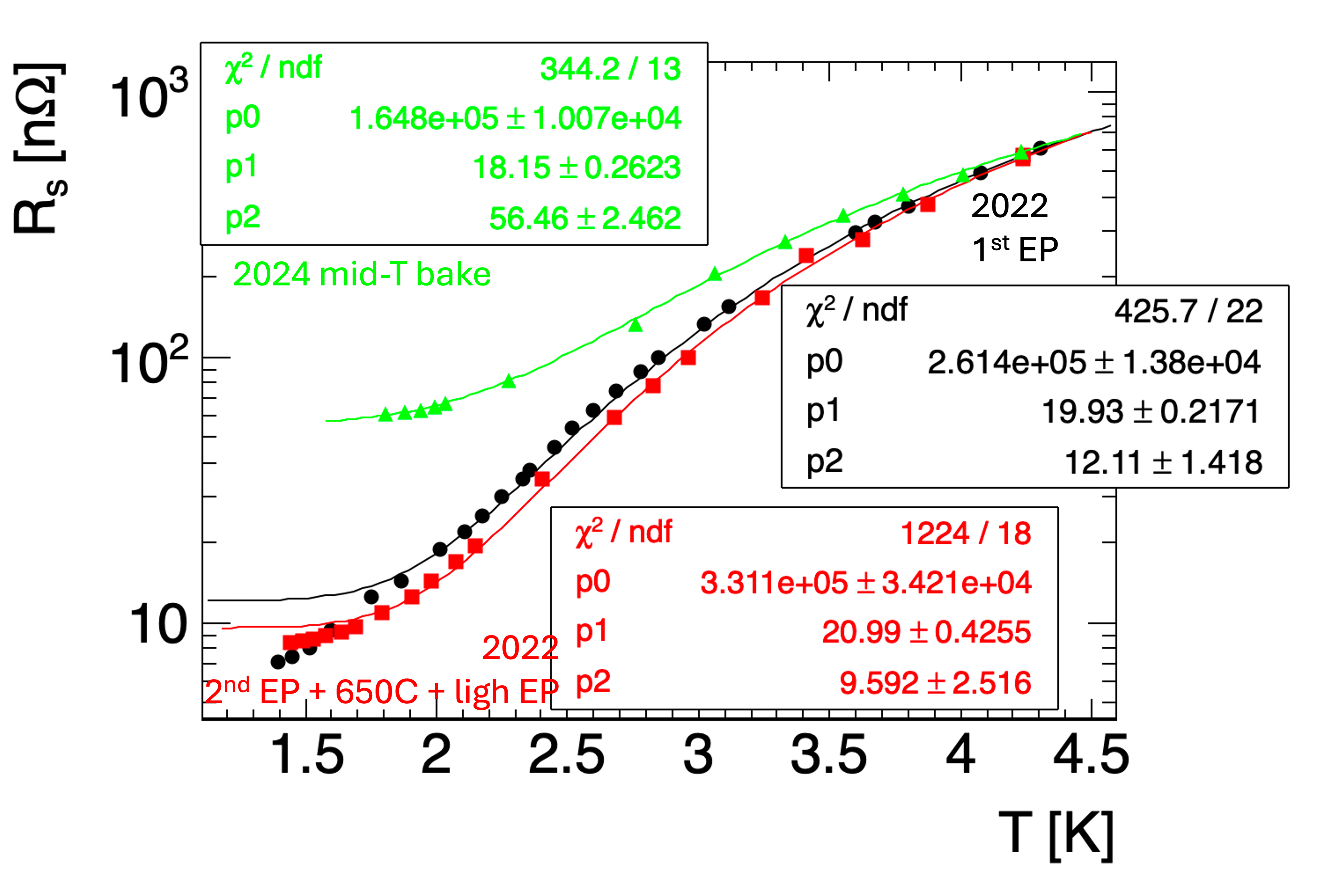}
\caption{Comparison of surface resistance as a function of temperature before and after mid-T bake. 
The black circle points show data after 1st EP. 
The red squares show data after 2nd EP and annealing at 650$^\circ$C for 10 hours followed by light EP. 
The green triangles show the data after mid-T bake newly presented in this article. 
The solid lines are fitting curves.}
\label{fig:Rs_vs_T}       
\end{figure}

After the heat treatment, the cavity was cleaned by high-pressure water rinsing, and vacuum flanges with RF antennas were attached.
The cavity was covered by a plastic bag and transported to CEA Saclay.
In June 2024, the cavity was mounted in a vertical test stand of CEA.
The cooling down from 300~K to 4~K was performed overnight.
We measured $Q_0$ at 1~MV/m during subatmospheric pumping from 4~K to below 2~K as shown in Fig.~\ref{fig:Rs_vs_T},
where surface resistance $R_{\rm s}$ is related to $Q_0$ via this formula:
\begin{equation}
R_{\rm s} = \frac{G}{Q_0},
\end{equation}
where $G=271\, \Omega$ is determined by the geometry of this cavity.
We observed clear degradation of surface resistance at cold compared to the results before the mid-T bake.

The solid lines in Fig.~\ref{fig:Rs_vs_T} show a fitting function with three parameters $(p0, p1, p2)$
\begin{equation}
R_{\rm s}(T) = R_{\rm BCS}(T) + R_{\rm res},
\end{equation}
with
\begin{equation}
R_{\rm BCS}(T) = \frac{p0}{T}\exp{\left(-\frac{p1}{T} \right)}
\end{equation}
and
\begin{equation}
R_{\rm res} = p2.
\end{equation}
The exponentially temperature-dependent term $R_{\rm BCS}$ is caused by thermally excited quasi-particles (normal conducting electrons).
The temperature-independent term $R_{\rm res}$ represents multiple effects including trapped magnetic flux.
The result clearly shows that the cavity after mid-T bake suffers from additional $\Delta R_{\rm res}\sim 46\, {\rm n\Omega}$.

\begin{figure}[h]
\centering
\includegraphics[width=10cm,clip]{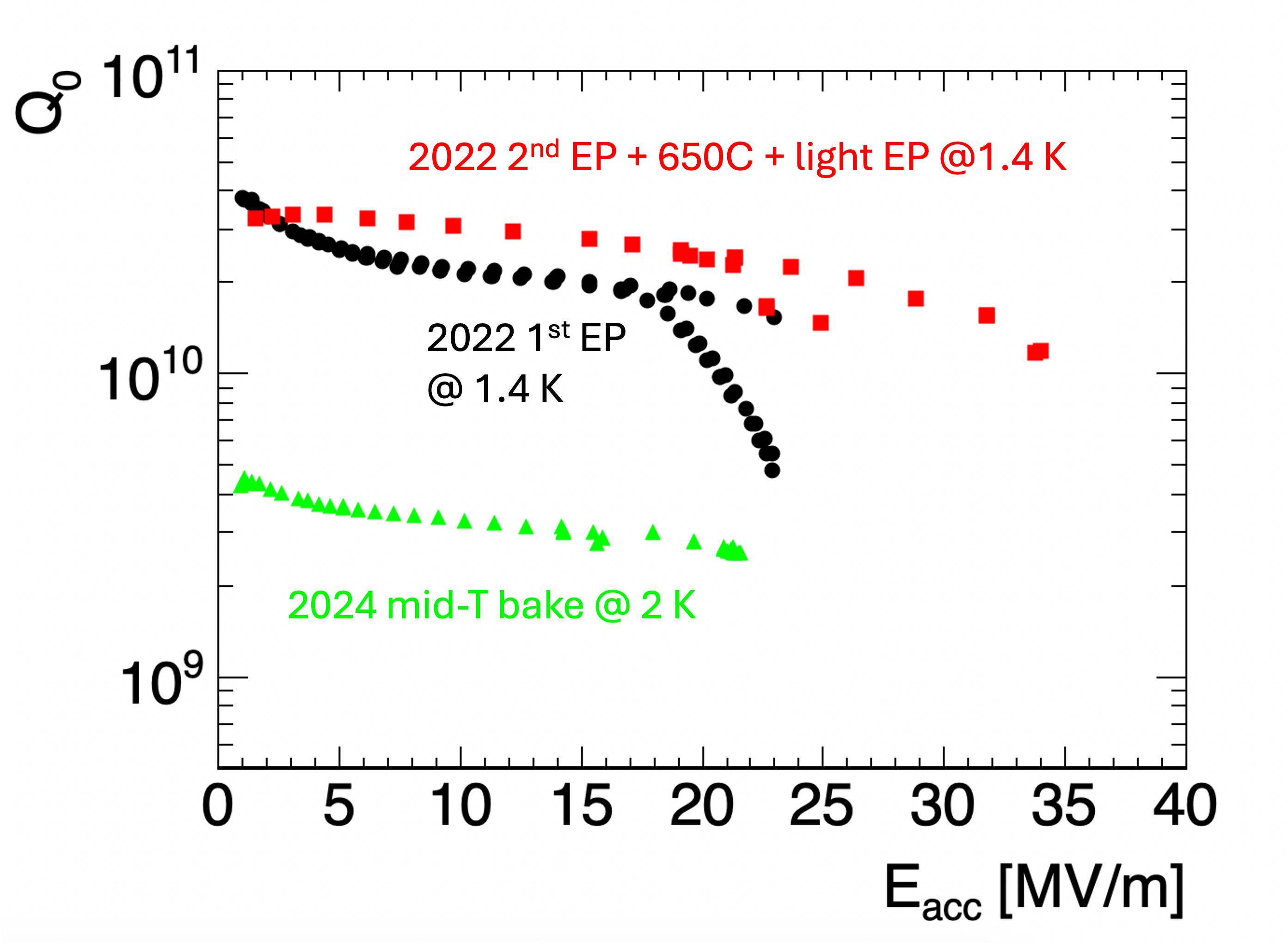}
\caption{Comparison of quality factor as a function of accelerating gradient before and after mid-T bake.
The black circle points show data after 1st EP. 
The red squares show data after 2nd EP and annealing and annealing at 650$^\circ$C for 10 hours followed by light EP. 
The green triangles show the data after mid-T bake newly presented in this article. 
}
\label{fig:Q_vs_E}       
\end{figure}

After the measurement during cooling down, 
we scanned the accelerating gradient at fixed temperature of 2~K as shown in Fig.~\ref{fig:Q_vs_E}.
Note that the previous data before mid-T bake was taken at 1.4~K.
This temperature 2~K was selected because it is the standard to compare with the literature of mid-T bake from other laboratories.
Despite the temperature difference, clearly, the cavity after mid-T bake shows significantly lower $Q_0$ and higher Q-slope (nonlinearly increasing surface resistance as a function of the field).

\begin{figure}[h]
\centering
\includegraphics[width=10cm,clip]{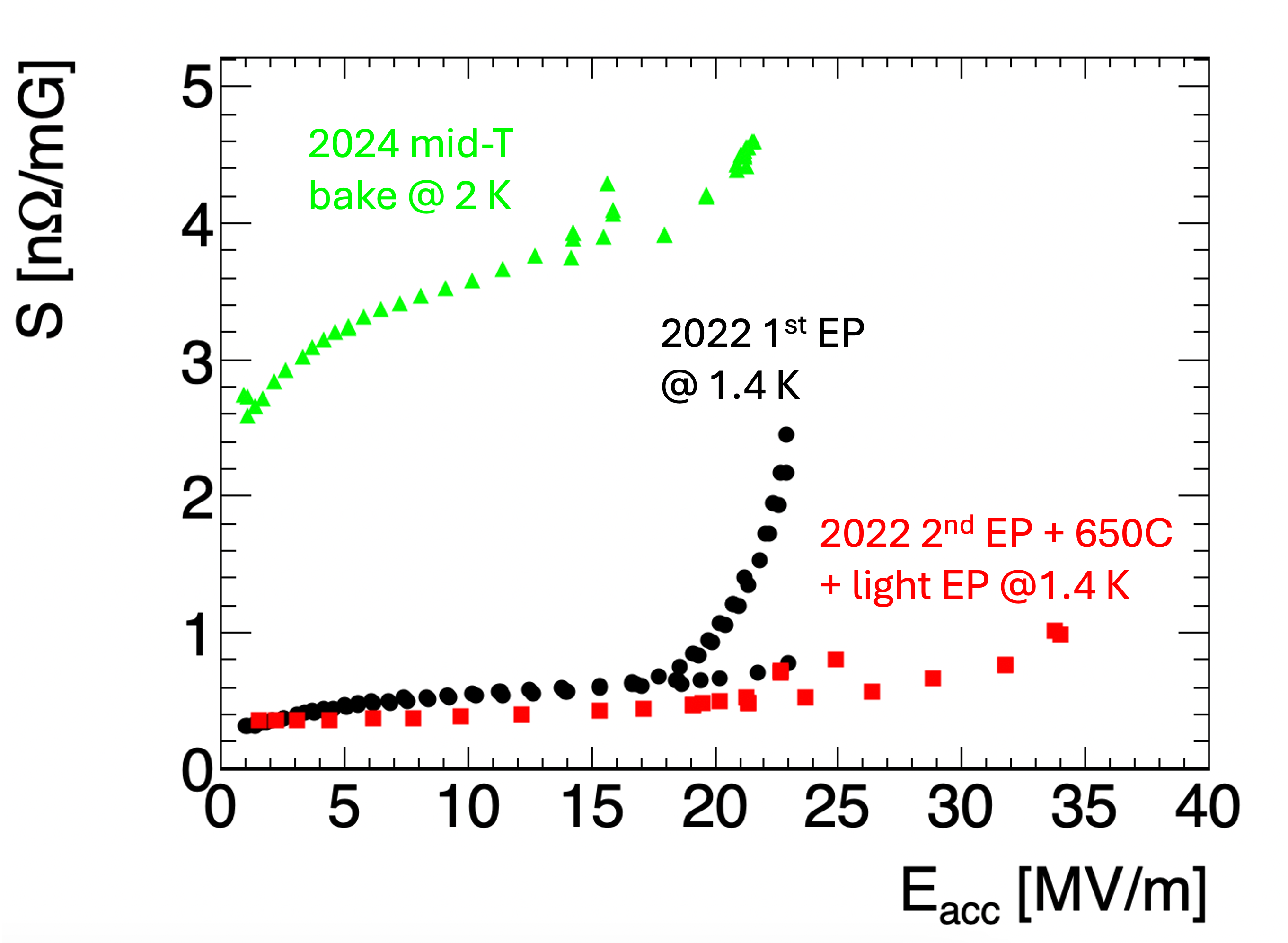}
\caption{Estimated flux sensitivity as a function of accelerating gradient, assuming trapped magnetic field of 25~mG.
The black circle points show data after 1st EP. 
The red squares show data after 2nd EP and annealing and annealing at 650$^\circ$C for 10 hours followed by light EP. 
The green triangles show the data after mid-T bake newly presented in this article. 
}
\label{fig:S_vs_E}       
\end{figure}

After the 1st cold test, we investigated residual magnetic fields inside the cryostat.
We observed around $H_{\rm ext}\sim$25~mG caused by unexpectedly magnetized components around the cavity.
This can cause trapped magnetic flux oscillating under the RF field and thus explain very high $R_{\rm res}$.
The surface resistance induced by such trapped flux can be expressed by a simple formula
\begin{equation}
R_{\rm fl} = S\times H_{\rm ext},
\end{equation}
with $S$ flux sensitivity and $H_{\rm ext}\sim 25$~mG without flux expulsion during cooling down.
For the data before mid-T bake, whose data were taken at 1.4~K, $R_{\rm BCS}$ is negligibly small; thus,
\begin{equation}
R_{\rm s}({\rm 1.4\, K}) \sim R_{\rm res} \sim R_{\rm fl}
\end{equation}
was supposed.
For the data after mid-T bake,  $R_{\rm s}$ may be dominated by $R_{\rm fl}$ even at 2~K.
Under these assumptions, Figure~\ref{fig:S_vs_E} was obtained.
The very high $S$ after mid-T bake is consistent with the previous studies~\cite{dhakal24impact, Ito:2022rok}.
Since the cavity became more sensitive to the magnetic field,
the cryostat and supporting structure of the cavity, 
which gave sufficiently low magnetic fields to the cavity before mid-T,
was no longer relevant to the measurement of this particular cavity.

\section{Future prospects} \label{sec:future}
In this study, we confirmed the very high sensitivity to magnetic fields after the mid-T baking process.
We identified the issue and will remove the magnetized components.
The 2nd test is planned to validate the cavity performance under the smaller residual magnetic field in July 2024.
Moreover, fast cooling down to expel the remaining magnetic field in the improved environment may be necessary to take comparable data to the literature.
This flux expulsion process was optimized for the LCLS-II project~\cite{PhysRevAccelBeams.22.032001}.
Since modification of the cryogenic system for fast cooling down takes some time,
we will organize the 3rd test of the same cavity in the cryostat at KEK in October 2024.
This cryostat has been used to qualify other mid-T baked cavities with excellent control in magnetic fields and cooling for flux expulsion~\cite{Ito:2022rok}.
We will compare this cavity with other reference cavities so that we can conclude the heat treatment process in our furnace.

\section{Conclusion and outlook} \label{sec:conclusion}
IJCLab is extending its research activities from low-$\beta$ cavities for heavy ion and proton accelerators to $\beta\sim1$ elliptical cavities for energy frontier colliders.
In particular, we are constructing a PERLE accelerator as a demonstrator of the ERL concept at a high current of 20~mA for more sustainable future accelerator projects in particle physics.
We identified that one of the key components is very high-$Q_0$ cavities with advanced heat treatment, namely, mid-T bake.
The mid-T baking process is matured in 1.3GHz and 650MHz cavities but its application to PERLE frequency range of 800MHz is novel.
We started to evaluate the process with an ILC-type cavity.
The 1st mid-T baking was performed in a clean vacuum furnace in IJCLab.
The 1st cold test revealed a very high sensitivity to the residual magnetic field after the mid-T bake,
which is consistent with the results from other laboratories.
We will test the cavity again with a better magnetic field environment and then with flux expulsion.

In parallel with the qualification of the mid-T baking process, 
we are developing 800MHz prototype cavities.
These cavities are PERLE prototype as well as one of the CERN-FCCee prototypes.
Once the furnace process is fully understood with the small ILC-type cavities,
we will eventually perform heat treatment with these relatively large 800MHz cavities.

\section*{Acknowledgement}
We thank Detlef Reschke and Lea Steder for useful discussions on the cavity originally from DESY.
Special thanks goes to Pashupati Dhakal, Kellen McGee, Jiyuan Zhai, and Peng Sha for their valuable inputs in an informal review meeting of this project.
This project has received funding from the European Union’s Horizon 2020 Research and Innovation programme under Grant Agreement No 101004730 and 101086276.
%
 \bibliography{SRF} 
%
%
%
%

\end{document}